\documentclass[cup5b]{caps}
\usepackage{graphicx}
\usepackage{amssymb}
\usepackage{ociwsymp4}

\def\aa{{A\&A}}
\def\aas{{A\&AS}}
\def\aj{{AJ}}
\def\annrev{{ARA\&A}}
\def\apj{{ApJ}}
\def\apjs{{ApJS}}

\def\mnras{{MNRAS}}

\def\pasp{{PASP}}
\newcommand\hbeta{\ensuremath{\mathrm{H}\beta}}
\newcommand\feh{\ensuremath{\mathrm{[Fe/H]}}}
\newcommand\afe{\ensuremath{\mathrm{[\alpha/Fe]}}}
\newcommand\mgb{\ensuremath{\mathrm{Mg}\,b}}
\newcommand\mgfe{\ensuremath{\mathrm{[MgFe]}}}
\newcommand\fe{\ensuremath{\mathrm{\langle Fe\rangle}}}
\newcommand{\z}{\ensuremath{\mathrm{[Z/H]}}}
\makeatletter
\newcommand\ion[2]{#1$\;${\small\rm\@Roman{#2}}\relax}
\makeatother

\HeadText{S. C. Trager}

\def\plotone#1{\centering \leavevmode
\includegraphics[width=.95\columnwidth]{#1}}

\def\plotone#1{\centering \leavevmode
\includegraphics[width=.95\columnwidth]{#1}}

\begin{document}
\pagenumbering{arabic}

\author[]{S. C. TRAGER\\Kapteyn Astronomical Institute,
Groningen, The Netherlands}

\chapter{Abundances from the Integrated \\
Light of Globular Clusters and \\
Galaxies}

\begin{abstract}
It is currently impossible to determine the abundances and ages of the
stellar populations of distant, dense stellar systems star by star.
Therefore, methods to analyze the composite light of stellar systems
are required.  I review the modelling and analysis of integrated
spectra of the stellar populations of individual globular clusters,
globular cluster systems, early-type galaxies, and the bulges of
spiral galaxies, with a focus on their abundances and abundance
ratios.  I conclude with a list of continuing difficulties in the
modelling that complicate the interpretation of integrated spectra as
well as a look ahead to new methods and new observations.
\end{abstract}

\section{Introduction}

The reviews presented in this volume demonstrate that the
nucleosynthetic history of the Milky Way and its satellites can now be
probed in exquisite detail, revealing a wealth of information about
the processes by which our own Galaxy and its neighbors have formed.
Such star-by-star analysis of the abundances of distant globular
clusters and distant and/or dense galaxies is however currently beyond
the reach of current telescopes.  Although this situation might
change with the advent of overwhelmingly large telescopes of the
30--100m class with high-precision adaptive optics systems, during the
20th Century and at the beginning of the 21st, we have been limited to
studying the \emph{integrated} stellar populations of distant globular
clusters and galaxies.

In this review, I will discuss spectroscopic techniques for
determining abundances from the integrated light of galaxies and
globular clusters.  I will begin with a discussion of the ingredients
of the models required to interpret the integrated-light spectra, the
calibration of those models, and a few of their their pitfalls.  I
will then discuss the abundances determined from integrated light
spectroscopy of globular clusters in our Galaxy, the Local Group, and
the globular cluster systems of early-type galaxies.  I will then
consider the abundances of nearby early-type galaxies and the bulges
of early-type spirals determined from integrated-light spectroscopy,
including recent results on abundance anomalies in these systems.
Although I will focus on deriving abundances from optical spectra,
these results necessarily depend on the \emph{ages} of the systems
(through the age--metallicity degeneracy, \S\ref{sec:tzdegen}), and
so I must briefly discuss the ages of these systems.  Finally, I will
summarize the current state of the observations, the current problems
in understanding stellar populations from integrated-light
spectroscopy, and current and future directions in this subject.

Perusal of this outline reveals that there are a number of topics I
will not discuss, due either to space constraints or to difficulties
in interpretation.  I will not be able to discuss the fascinating
history of this subject in detail.  I will not discuss purely
photometric methods of determining abundances, such as attempts to
derive abundances of globular clusters and early-type galaxies from
broad-band colors, as the interpretation of broad-band optical colors
is seriously compromised by the age--metallicity degeneracy.  I will
not have space to discuss abundances derived from wavebands other than
the optical; there have been excellent recent reviews and studies of
the stellar populations of globular clusters and early-type galaxies
determined from, e.g., UV spectroscopy (O'Connell 1999; Dorman,
O'Connell \& Rood 2003; Peterson et al.\ 2003; Peterson, this
meeting).  Last, but not least, I will not discuss stellar abundances
of irregular or dwarf galaxies themselves (apart from M32), as this
subject is ably covered in Section V of this volume, and I will touch
only lightly on abundances derived from the integrated \emph{stellar}
light of bulges of early-type spiral galaxies.  Garnett (this volume)
thoroughly and cogently discusses the \emph{gas-phase} abundances of
such galaxies.

\section{Stellar population modeling of integrated-light spectra}
\label{sec:models}

The interpretation of the integrated light of globular clusters and
galaxies requires astrophysically-constrained models.  In this
section, I outline the basic purpose and ingredients of such models
and the steps required to calibrate them.

\subsection{The basic problem}
\label{sec:tzdegen}

\begin{figure}
\plotone{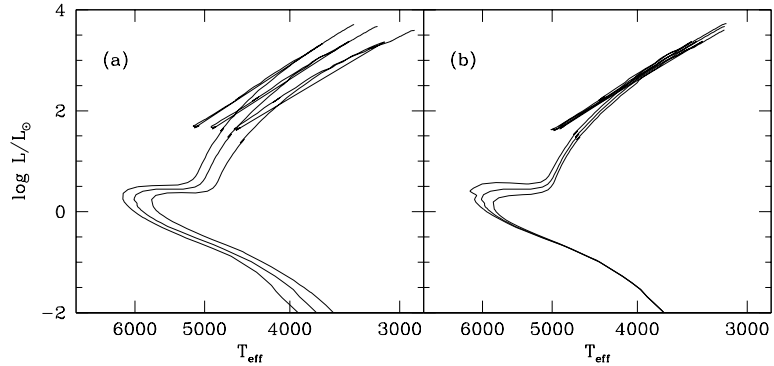}
\caption{The age--metallicity degeneracy.  (a) Varying metallicity at
  fixed age: 12 Gyr old Padova (Girardi et al.\ 2000) isochrones with
  metallicities of $\feh=-0.68$, $-0.38$, 0 (left to right).  (b)
  Varying age at fixed metallicity: $\feh=-0.38$ Padova isochrones
  with ages of $t=7$, 10, 14 Gyr (left to right).  Note that in this
  view of the age--metallicity degeneracy, the main sequence turnoff
  (MSTO) varies in temperature roughly equally in each case, but the
  red giant branch (RGB) varies little with age at fixed metallicity.
  To break the degeneracy, one needs to measure the temperature of
  both the MSTO and the RGB to decouple the age and
  metallicity.\label{fig:tzdegen}}
\end{figure}

O'Connell (1986) gives a clear physical explanation of the
\emph{age--metallicity degeneracy} that plagues the study of the
integrated light of stellar populations.  Because the emergent flux
from a stellar atmosphere is a superposition of spectra which are
characterized by different temperatures weighted by a function of
opacity, increases in opacity in a stellar atmosphere due to increased
metallicity are equivalent to a decreased mean temperature for the
emergent flux.  Therefore, a metal-rich population can be simulated by
decreasing the mean temperature---that is, decreasing the age---of a
more metal-poor population (Fig.~\ref{fig:tzdegen}).

The key breakthrough in the analysis of integrated stellar populations
was made by Rabin (1980, 1982) and Gunn, Stryker \& Tinsley (1981),
who noticed independently that the strengths of the Balmer lines of
hydrogens derived from evolutionary population synthesis models allow
an accurate age of a single-burst stellar population to be measured.
Rabin expressed this idea in a powerful graphical form: the
``hydrogen--metals diagnostic diagram,'' in which the strength of
Balmers are plotted as a function of a metal-line (in this case,
\ion{Ca}{2} K) for the integrated spectra and stellar population
models.  Using this diagram, he could break the age--metallicity
degeneracy.  Applying this method to clusters in the Magellanic
Clouds, Rabin confirmed that the Searle, Wilkinson \& Bagnuolo (1980;
hereafter SWB) ranking of those clusters was indeed an age ranking.

Worthey (1994) and Worthey \& Ottaviani (1997) quantified the
age--metallicity degeneracy for a large number of absorption-line
strengths and broad-band colors.  Metal lines (like \mgb\ and \fe) and
broad-band colors are slightly more sensitive to metallicity than age,
in the ratio $\partial\log t/\partial \log Z=3/2$ (Worthey 1994;
Fig.~\ref{fig:tzdegen}).  This means that a factor of 3 change in age
looks like a factor of 2 change in metallicity for a simple stellar
population.  Clearly, age-sensitive indices are desired---lines that
preferentially measure the (luminosity-weighted) temperature of the
main-sequence turnoff.  Following Rabin (1980, 1982), Worthey (1994)
showed that $\partial\log t/\partial \log Z\approx2/3$ for \hbeta\
(and $\approx1$ for H$\delta$ and H$\gamma$; Worthey \& Ottaviani
1997).  Thus, a Balmer-line index \emph{combined} with a metal-line
index \emph{breaks} the age--metallicity degeneracy.

\subsection{Ingredients}
\label{sec:ingredients}

The first attempt to synthesize the spectral lines and colors of a
galaxy appears to have been made by Whipple (1935).  Whipple developed
population synthesis, in which arbitrary permutations of stellar types
are combined to make a synthetic galaxy spectrum and colors; this
method was highly influential on many later workers (e.g., de
Vaucouleurs \& de Vaucouleurs 1959; Spinrad 1962a,b; Spinrad \& Taylor
1969, 1971; Lasker 1970; Faber 1972; O'Connell 1976; Williams 1976).
Tinsley (1968, 1972; Tinsley \& Gunn 1976) invented the method that
has generally superseded population synthesis: \emph{evolutionary}
population synthesis.  In this method, populations are modelled from
the starting point of an isochrone and a luminosity function and then
matched to the observations, instead of attempting to extract a
color-magnitude diagram and luminosity function from the observations.
Tinsley's method has become the basis for modern stellar population
models (e.g., Bruzual 1983; Buzzoni 1989; Charlot \& Bruzual 1991;
Bruzual \& Charlot 1993; Worthey 1994; Bressan, Chiosi \& Fagotto
1994; Maraston 1998).

To build a model of a \emph{simple stellar population} (SSP)---that
is, a single-age, single-metallicity stellar population---that can be
used to analyze the integrated spectrum of a given object, four major
ingredients are required.
\begin{description}
\item[Isochrones] An isochrone set that is best calibrated against the
populations of interest (young, intermediate-aged, or old; metal-poor
or metal-rich) is needed as the basic astrophysical constraint.
Popular choices for the intermediate-aged and old populations
considered here are currently the Padova set (Girardi et al.\ 2000;
Salasnich et al.\ 2000); the set from Cassisi, Castellani \&
Castellani (1997) and Bono et al.\ (1997); and the set of Salaris \&
Weiss (1998) and more recent extensions (see, e.g., Schiavon et al.\
2002b).
\item[Initial mass function] An IMF populates the isochrones as
needed.  Typically, a Salpeter (1955) IMF is chosen, although other
choices are possible (Vazdekis et al.\ 1996).
\item[Stellar fluxes] While it may be preferable to have a set of
observed stellar fluxes covering a comprehensive range in temperature,
gravity and metallicity (and even \afe), such a set is not currently
available (although see Vazdekis 1999 for another approach and Le
Borgne et al.\ 2003 for a hint of what is to come).  We therefore are
reliant on the theoretical flux library of Kurucz (1993), which covers
a broad range of parameter space, or modifications of that library to
attempt to bring its colors into better agreement with real stars
(Lejeune, Cuisinier \& Buser 1997, 1998; Westera et al.\ 2002).
\item[Stellar spectra or absorption-line strengths] Because we require
the strengths of Balmer-line and metal-line indices to break the
age--metallicity degeneracy, a library of either stellar spectra
(e.g., Jones 1998) or stellar absorption-line strengths (e.g., Worthey
et al.\ 1994) is required.  Given the difficulties matching the
Lick/IDS system precisely (Worthey \& Ottaviani 1997; Trager, Faber \&
Dressler 2003), we would like to find new libraries for population
syntheses.  Such libraries are on their way (Schiavon, priv.~comm.;
Worthey, priv.~comm.; Peletier, priv.~comm.; Rose, priv.~comm.).
\end{description}

\begin{figure}
\plotone{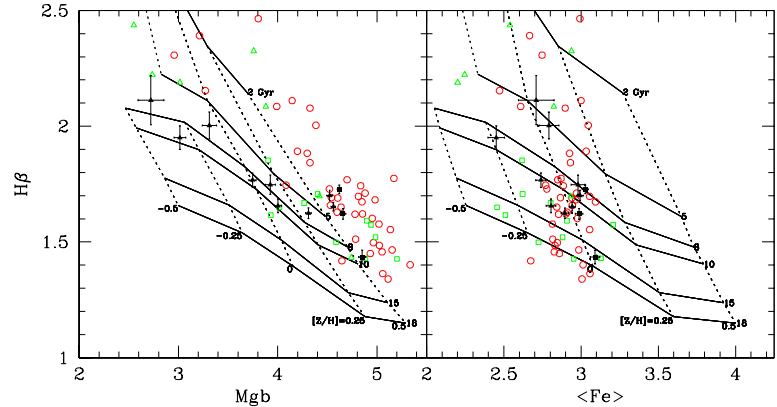}
\caption{The line strengths of local early-type galaxies compared with
  stellar population models from Worthey (1994).  Solid lines:
  constant age.  Dashed lines: constant metallicity.  Solid squares
  and triangles: elliptical and S0 galaxies in Coma (Trager, Faber \&
  Dressler 2003).  Open squares and triangles: elliptical and S0
  galaxies in Fornax (Kuntschner 2000).  Open circles: elliptical
  galaxies in Virgo, groups, and the field (Gonz\'alez 1993).  Left:
  \hbeta\ as a function of \mgb.  Right: \hbeta\ as a function of \fe.
  Note the significant difference in derived age and metallicity for
  most galaxies from each pair of indices; this is indicative of
  $\afe\neq0$ for these galaxies. \label{fig:worthey}}
\end{figure}

Once these ingredients are available, one can construct a stellar
population model, as shown in Figure~\ref{fig:worthey}.  However, such
models clearly have a problem: ages and metallicities derived from
different pairs of Balmer- and metal-line indices are different for
giant elliptical and many S0 galaxies.  These discrepancies result
from the non-solar abundance ratios of these objects (O'Connell 1976;
Faber \& Jackson 1976; Peterson 1976; Peletier 1989; Worthey, Faber \&
Gonz\'alez 1992; Greggio 1997; Worthey 1998; Trager et al.\ 2000a).
Although we refer to these non-solar abundance ratios by the shorthand
phrase ``$\alpha$-enhancements,'' we stress that the proper
interpretation of these discrepancies is actually as
``Fe-deficiencies,'' because the ``$\alpha$-elements'' O and Mg
dominate the metallicity determinations (Greggio 1997; Trager et al.\
2000a).  Correcting for $\afe\neq0$ (or more generally, for
$[X_i/\mathrm{Fe}]\neq0$ for any element $X_i$: TMB03) requires either
an empirical stellar library with the needed enhancements or stellar
atmosphere models in which synthetic spectra with the needed
enhancements can be generated, as well as isochrones with $\afe\neq0$,
such as those available in the Padova and Salaris \& Weiss sets (cf.\
Trager et al.\ 2000a; but see Maraston \& Thomas 2003).  The empirical
approach was attempted first by Weiss, Peletier \& Matteucci (1995),
who coupled fitting functions for Mg$_2$ and \fe\ from the Bulge
giants of Rich (1988) with $\alpha$-enhanced isochrones;
unfortunately, the Rich library was not complete enough for a
reasonable coverage of parameter space and more importantly did not
include the crucial \hbeta\ index.  Trager et al.\ (2000a) and TMB03
have chosen to use the stellar atmosphere computations of Tripicco \&
Bell (1995), who modelled the response of the Lick/IDS index system to
changes in individual elements, in a differential sense, and ignore
the $\alpha$-enhanced isochrones (Maraston \& Thomas 2003).  While
more flexible for analyzing populations that may not resemble local
stars, calibration of these theoretical corrections is required.

\subsection{Calibrations}
\label{sec:calib}

\begin{figure}
\plotone{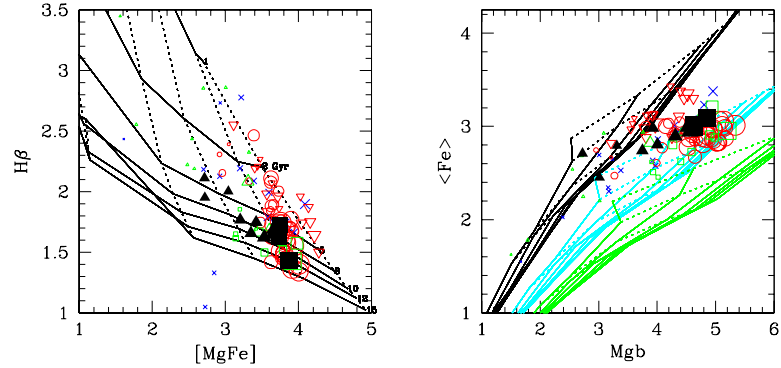}
\caption{The line strengths of the nuclei of local early-type galaxies
  compared with stellar population models from TMB03.  Points and line
  types as in Fig.~\ref{fig:worthey}, except inverted triangles: S0's
  in groups and Virgo (Fisher, Franx \& Illingworth 1996) and crosses:
  bulges of S0/a--Sbc spirals in various environments (Proctor \&
  Sansom 2002).  Point sizes for all samples are proportional to
  velocity dispersion.  Left: Ages and metallicities.  Note that the
  metallicity (left to right: $\feh=-1.35$, $-0.33$, 0, $0.35$, and
  $0.67$ dex) and age scales have changed.  Note that field
  ellipticals (circles), nearly all S0's, and bulges span a large
  range in \hbeta\ strength and thus in SSP age, while cluster
  ellipticals (squares) are quite old and vary more in metallicity.
  Right: \afe\ (from left to right: $\afe=0$, $+0.3$, $+0.5$ dex). In
  this plot, note that ellipticals tend to have $\afe\approx+0.3$---in
  fact, this varies with velocity dispersion, with small ellipticals
  being more ``solar'' and large ones being more enhanced (Trager et
  al.\ 2000b)---while S0's and bulges tend to be more ``solar.''
  \label{fig:hbmgfe_all}}
\end{figure}

I show the results of $\alpha$-enhanced models (TMB03) in
Figure~\ref{fig:hbmgfe_all}.  It is clear that if these models are
indeed properly calibrated, then it is possible to determine uniquely
age, metallicity, and \afe\ for a single-burst population.  Galaxies
are unlikely to be SSP's, but globular clusters are (except maybe
$\omega$ Cen: Smith, this volume).  I note here that the
``metallicities'' measured by the Trager et al.\ (2000a) technique are
total metallicities \z, not \feh.  However, \z\ can be converted to
\feh\ on the Zinn \& West (1984) scale with the following
(model-dependent) scaling: $\feh = \z - 0.94\afe$ (Tantalo, Chiosi \&
Bressan 1998; Trager et al.\ 2000a; TMB03).

I will now exploit this fact and turn to a model of an individual
cluster: 47 Tuc.  Gibson et al.\ (1999) pointed out that the stellar
population models of Jones \& Worthey (1995) predicted an age in
excess of 20 Gyr from H$\gamma$ and the Ca4227 or C$_2$4668 indices.
for 47 Tuc, a cluster with a CMD-based age of 11--14 Gyr (depending on
the isochrone set used; Schiavon et al.\ 2002b).  Vazdekis et al.\
(2001) brought the spectroscopic age down by using $\alpha$-enhanced
isochrones with He diffusion (from the Salaris \& Weiss 1998 set).
Schiavon et al.\ (2002a,b) used a two-pronged approach to solve this
problem.  First, they synthesized the spectrum directly from the CMD
(Schiavon et al.\ 2002a), which was meant to reveal any problems with
the stellar libraries.  Schiavon et al.\ uncovered several: (a) it is
necessary to have a strictly \emph{homogeneous} set of atmospheric
parameters for the stars in the library; (b) the metallicities of the
library stars used in the synthesis must be that of the cluster to
less than $0.05$ dex, and must be on the same metallicity scale; and
(c) the line strengths need to account for the CN bimodality of the
cluster (which extends to at least the main-sequence turnoff: Hesser
1978), in which roughly half the stars are CN-strong.  Now that the
necessary corrections to the line strengths are understood, the second
step is to synthesize the line strengths from the best-fitting
isochrone (Schiavon et al.\ 2002b).  A significant problem arises in
this step: the observed luminosity function does not match that of the
model, due to a lack of AGB stars (in the Salaris isochrones) and a
strong deficit of RGB stars (in both the Salaris and Padova
isochrones).  Once all of these problems are corrected for, nearly all
of the synthesized spectral indices match the observations at the
correct age (11--13 or 12--14 Gyr using the Salaris and Padova
isochrones, respectively) and metallicity ($\feh=-0.75$ dex).

Now that we have confidence in our stellar population models to
predict the age and metallicity of a single globular cluster (finally;
Searle 1986), we turn to the calibration of the corrections for
$\afe\neq0$.  Maraston et al.\ (2003) have done this by comparing the
Lick/IDS globular cluster data and the Bulge cluster data of Puzia et
al.\ (2002) with the TMB03 stellar population models, index-by-index.
I show Puzia et al.'s data for a few indices below
(Fig.~\ref{fig:mwm31gcs}) but comment here that several indices do not
match the models: CN$_1$ and CN$_2$ (Burstein et al.\ 1984); Ca4455
(this difficult index probably should be dropped from the Lick/IDS
system); strikingly, C$_2$4668, an index with a large dynamic range at
fixed temperature but varying metallicity (Worthey et al 1994; Worthey
1994); Na D; and the TiO indices (although this isn't surprising, as
Tripicco \& Bell 1995 did not model TiO properly).  TMB03 have solved
the discrepancies in the CN indices and C$_2$4668 by allowing
$\mathrm{[N/\alpha]}=+0.5$ (cf.\ Brodie \& Huchra 1991; Smith et al.\
1997); but Na D remains a mystery (Spinrad 1962a,b; Cohen et al.\
2003).

\section{Globular cluster systems}
\label{sec:gcs}

Now that the required tools are in hand, I turn to what has been
learned about the abundances and ages of the stellar populations of
globular cluster systems of the Milky Way and M31
(\S\ref{sec:mwm31gcs}), dwarf galaxies in the Local Group
(\S\ref{sec:dwarfgcs}), and early-type galaxies
(\S\ref{sec:exgalgcs}).

\subsection{The globular cluster systems of the Milky Way and Messier 31}
\label{sec:mwm31gcs}

Van den Bergh's (1969) spectroscopy of globular clusters in M31, the
Fornax dwarf spheroidal galaxy and the Milky Way opened the field of
extragalactic globular cluster systems.  He found that the
metallicities of M31 globular clusters were on average more metal-rich
than those in the Milky Way with a similar or slightly larger
metallicity spread, but the Fornax clusters were more metal-poor and
had a much smaller metallicity spread.  Interestingly, van den Bergh
also found that the Balmer line strengths of globular clusters may
differ at fixed metallicity, even in the most metal-rich clusters (see
the discussion in Rose 1985); this suggests that age may play a role
in the stellar populations of some globular cluster systems or that
blue horizontal branches may exist at much higher metallicities than
seen in the Milky Way globular clusters.

\begin{figure}
\plotone{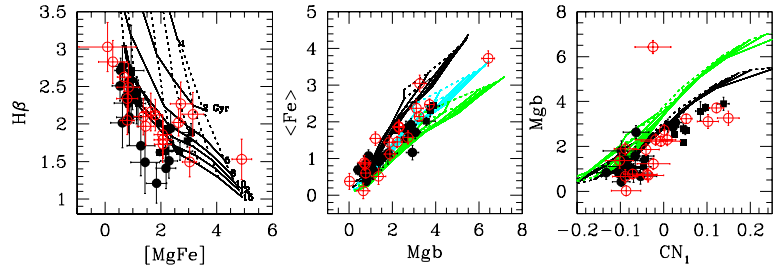}
\caption{The stellar populations of Milky Way (solid circles: halo
clusters from Trager et al.\ 1998; solid squares: Bulge clusters from
Puzia et al.\ 2002) and M31 (Trager et al.\ 1998; halo clusters only)
globular clusters.  Left: ages and metallicities.  Grids as in
Fig.~\ref{fig:hbmgfe_all}, although here extended down to $\feh=-2.25$
dex.  The two \hbeta-strong M31 clusters (V87, V116) are likely to be
intermediate-aged, but the slightly ``younger'' Milky Way clusters are
the anomalous BHB clusters NGC 6388 and NGC 6441 (Rose \& Tripicco
1986; Rich et al.\ 1997), which are as old as the other Galactic
clusters.  Center: \afe\ ratios; grids as in
Fig.~\ref{fig:hbmgfe_all}.  Right: The ``CN anomaly'' (Burstein et
al.\ 1984) in the Galactic and M31 clusters; grids are $\afe=0$
(lower) and $\afe=+0.5$ (upper) from TMB03.  Clearly, the Galactic and
M31 clusters do not differ significantly in \mgb/CN$_1$, but as a
whole are too strong in CN$_1$ at fixed \mgb\ given their typical
\afe\ (middle panel; Maraston et al.\ 2003; TMB03).  The highly
aberrant M31 ``cluster'' V204 is likely to be one of the brightest
stars in M31 (Berkhuijsen et al.\ 1988) rather than a globular
cluster.\label{fig:mwm31gcs}}
\end{figure}

In Figure~\ref{fig:mwm31gcs} I plot the line strengths of globular
clusters in the Milky Way and M31.  Three primary results can be read
from this Figure.
\begin{enumerate}
\item The mean metallicity and spread in metallicity of the globular
cluster systems in the two dominant spirals of the Local Group are
nearly identical when the bulge clusters of the Milky Way are
included.\footnote{Figure~\ref{fig:mwm31gcs} ignores the inner bulge
clusters of M31, for which spectral indices are unavailable.  It may
be that those clusters have high metallicity (as the Milky bulge
globulars have higher metallicity than the halo clusters), which would
weaken this conclusion.}
\item The M31 clusters are slightly more ``solar'' in their \afe\
ratios at the highest metallicities.
\item The CN strengths at fixed \mgb\ are very
similar in both systems but are significantly higher than the strength
predicted by the models at the \afe\ ratios derived from the
\mgb--\fe\ diagram.  
\end{enumerate}
Point (1) implies that metallicities of M31 clusters from
integrated-light spectra interpreted using the TMB03 models can be
assumed to be well-calibrated onto the Zinn \& West (1984) scale as
long as \afe\ can be measured.  Point (2) supports the identification
of the two ``intermediate-aged'' (\hbeta-strong) M31 clusters (V87 and
V116) as being somewhat younger than the dominant cluster populations
in the two systems (cf.\ Brodie \& Huchra 1990).  However, the
\hbeta-strong, metal-rich Galactic globular clusters are the Bulge
clusters NGC 6388 and NGC 6441, which have well-populated blue
horizontal branches that do not exist in other metal-rich clusters
(Rose \& Tripicco 1986; Rich et al.\ 1997).  Point (3) disagrees with
the conclusions of Burstein et al.\ (1984), who found that M31
clusters have stronger CN strengths than Galactic globular clusters,
solely because the Lick/IDS database does not contain the metal-rich
Bulge clusters which span the high-CN strengths of the metal-rich M31
clusters.  As discussed above, ``enhanced'' CN at fixed \mgb\ appears
to be a result of $\mathrm{[N/\alpha]}=+0.5$ in these clusters.

I note however that many of the globular clusters in
Figure~\ref{fig:mwm31gcs} fall below the \hbeta\ strengths of the
TMB03 models.  As discussed in \S\ref{sec:calib} above, this is due to
a number of factors.  TMB03 use the Worthey et al.\ (1994) fitting
functions, taken from the Lick/IDS stellar library of absorption-line
strengths, in which the atmospheric parameters are somewhat
inhomogeneous, and which are constrained by very few stars at
metallicities below $\feh<-1.25$, which makes direct comparison with
halo globular clusters problematic.  Finally, it is possible that the
TMB03 models may have problems with the RGB/AGB luminosity functions,
although the Maraston (1998) synthesis method on which TMB03 is based
uses the fuel-consumption theorem (Renzini \& Buzzoni 1986), which
should properly populate the luminosity functions.

\subsection{The globular cluster systems of dwarf galaxies in the
Local Group}
\label{sec:dwarfgcs}

\begin{figure}
\plotone{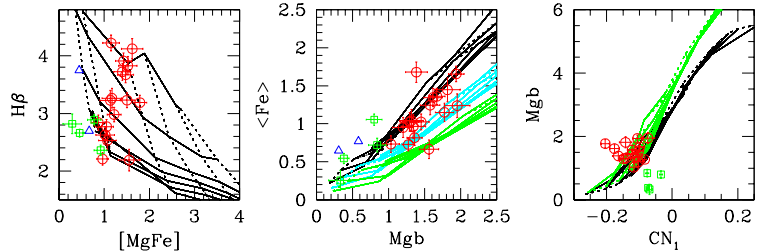}
\caption{The stellar populations of globular cluster systems of Local
Group dwarf galaxies.  Circles: LMC clusters from Beasley, Hoyle \&
Sharples (2002; for SWB classes IVB--VII only).  Squares: Fornax dSph
clusters from Strader et al.\ (2003a).  Triangles: NGC 6822 clusters
from Cohen \& Blakeslee (1998; errors were not published).  Panels and
models as in Fig.~\ref{fig:mwm31gcs}.  Note the CN anomaly in the
Fornax dSph globular cluster system. \label{fig:dwarfgcs}}
\end{figure}

In Figure~\ref{fig:dwarfgcs} I show the stellar populations for the
globular cluster systems of three dwarf galaxies in the Local Group:
the Large Magellanic Cloud (Beasley, Hoyle \& Sharples 2002), the
Fornax dSph (Strader et al.\ 2003b), and the dIrr NGC 6822 (Cohen \&
Blakeslee 1998; see also Chandar, Bianchi \& Ford 2000; Strader,
Brodie \& Huchra 2003b).  As shown previously by Rabin (1982), Beasley
et al.\ (2002) find that the SWB sequence is indeed an age sequence,
with the younger clusters being more metal-rich.  In fact, the LMC
clusters represent the best training set we currently have for
calibrating the stellar populations of intermediate-aged populations
(Searle 1986), but the multivalued nature of the TMB03 model grids at
very low metallicity (because of the blue horizontal branches) makes
the old, metal-poor end difficult to calibrate without \emph{a priori}
knowledge of the horizontal branch morphology (Beasley et al.\ 2002).
The \afe\ ratios of the LMC clusters---which is only slightly
super-solar---do not seem to decrease appreciably with decreasing age,
so possibly age and \afe\ are not tightly coupled (as is also seen for
giant ellipticals, Trager et al.\ 2000b; Fig.~\ref{fig:hbmgfe_all}).
Moreover, the LMC clusters appear not to have a significant CN
anomaly.  The reasons for this are unclear, but may be related to the
$\sigma$--\afe\ relation discussed below for early-type galaxies.

The two smaller dwarfs considered in Figure~\ref{fig:dwarfgcs} have
very metal-poor globular cluster populations.  The clusters in Fornax
are quite old and even show a CN anomaly of the sort present in the
Galactic and M31 clusters---that is, the CN strengths are too high at
fixed \mgb\ strength.  The NGC 6822 clusters have a range of ages,
with at least one old cluster (H VII) and a number of younger clusters
(Cohen \& Blakeslee 1998; Chandar et al.\ 2000; Strader et al.\
2000b).  Da Costa \& Mould (1988) employed a method similar to Rabin's
(1982) study to probe the stellar populations of the globular cluster
systems of NGC 147, NGC 185, and NGC 205, comparing the equivalent
widths of the Balmer lines with \ion{Ca}{2}\ K.  They found that the
majority of the clusters were old, although both NGC 205 and NGC 185
have intermediate-aged clusters and NGC 205 has at least two
metal-rich clusters similar to those in the Galactic Bulge (see their
Fig.~3).  However, line strengths on the Lick/IDS system do not exist
for these clusters and direct comparison with the stellar populations
derived from Figure~\ref{fig:dwarfgcs} is not possible.

\subsection{The globular cluster systems of nearby early-type
galaxies}
\label{sec:exgalgcs}

Although heroic early efforts were made by Racine, Oke \& Searle
(1978), Hanes \& Brodie (1986), Mould, Oke \& Nemec (1987), Mould et
al.\ (1990), Brodie \& Huchra (1991), Perelmuter, Brodie \& Huchra
(1995), Jablonka et al.\ (1996), Minniti et al.\ (1996), and Bridges
et al.\ (1997) using 4--5m class telescopes, the era of measuring
accurate stellar populations of the globular cluster systems of
early-type galaxies began in earnest with the availability of
multislit spectrographs on 8--10m class telescopes.  Kissler-Patig et
al.\ (1998) and Cohen, Blakeslee \& Rhyzov (1998) studied NGC 1399 and
M87, respectively, using the LRIS multislit spectrograph (Oke et al.\
1995) on the Keck Telescopes.  Cohen et al.\ (1998) studied the
largest sample of extragalactic globular clusters around a single
galaxy to date (150).  Although the typical signal-to-noise of these
spectra is not as good as the most recent studies (e.g., Puzia et al.\
2003), given the size of the dataset, statistically significant
conclusions about the stellar populations of the globular cluster
system of an early-type galaxy could be drawn for the first time.  In
particular, $\approx95\%$ of the clusters around M87 are old and can
be classified into metal-poor ($\feh\approx-1.3$) blue clusters and
metal-rich ($\feh\approx-0.7$) red clusters in a ratio of 2:3, thus
resolving most of the debate about the cause of the bimodality in
colors of globular cluster systems of early-type galaxies (e.g., Kundu
\& Whitmore 2001; Larsen et al.\ 2001).  However, a small
($\approx5\%$) population of intermediate-aged clusters can also be
found in M87.

\begin{figure}
\plotone{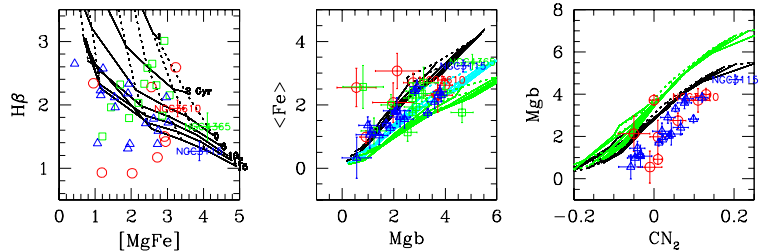}
\caption{The stellar populations of the globular cluster systems of
nearby early-type galaxies.  Panels and models as in
Fig.~\ref{fig:mwm31gcs}.  Triangles: NGC 3115 (Kuntschner et al.\
2002).  Circles: NGC 3610 (Strader et al.\ 2003c).  Squares: NGC 4365
(Larsen et al.\ 2003; no CN data available).  Host galaxies are
labelled with their ID; data taken from Fisher et al.\ (1996; NGC
3115) or Trager et al.\ (1998; NGC 3610, NGC 4365).  Cluster
line-strength errors in the leftmost panel have been removed for
clarity; typical errors are
$\sigma_{\hbeta}\approx\sigma_{\mgfe}\approx0.5$ \AA\ (slightly
smaller for NGC 3115).  Note that the cluster systems of NGC 3115 and
NGC 3610 apparently have a CN anomaly. \label{fig:exgalgcs}}
\end{figure}

In Figure~\ref{fig:exgalgcs} I show recent results for the globular
cluster systems of three early-type galaxies: NGC 3115 (Kuntschner et
al.\ 2002); NGC 3610 (Strader et al.\ 2003c); and NGC 4365 (Larson et
al.\ 2003), together with line strengths of their host galaxies (see
figure caption for line-strength references).  Although the error bars
are large, the basic results of Cohen et al.\ (1998) are confirmed:
there are generally three populations of globular clusters in
early-type galaxies---metal-poor, old, blue clusters, metal-rich, old,
red clusters, and often small frostings of metal-rich,
intermediate-aged, red(der) clusters (see also Puzia et al.\ 2003).
This small frosting of intermediate-aged clusters may be a key to
understanding the formation history of their host galaxies.  A
particularly interesting case is the globular cluster system of NGC
4365, as NGC 4365 has no evidence of having an intermediate-aged
stellar population (Davies et al.\ 2001; Larsen et al.\ 2003;
Fig.~\ref{fig:exgalgcs}).  It may be hoped that the
multiple-population degeneracy of the Lick/IDS index system (Trager et
al.\ 2000b) can be broken using globular clusters as tracers of
individual star formation events in their host galaxies; such work is
in progress (e.g., Larsen et al.\ 2003).  Note also that these
globular cluster systems also show a CN anomaly.

\section{Early-type galaxies and bulges}

Determining the abundances and ages of galaxies is far more complex
than doing the same for globular clusters, which are nearly all SSPs.
Galaxies are composite stellar populations, and thus the
interpretation of their spectra suffer from (currently) unavoidable
difficulties.  In particular, the absorption-line strength indices of
composite populations are the luminosity-weighted sums of their SSP
components.  This means that young populations, which have low
mass-to-light ratios and therefore high light-to-mass ratios, unduly
bias the analysis of composite populations if present (see, e.g.,
Trager et al.\ 2000b).  However, in the absence of other knowledge
(like the ages of the globular clusters possessed by the galaxy),
line-strength indices only give the \emph{SSP-equivalent} stellar
population parameters for any individual galaxy.  Therefore, I will
proceed with a discussion of the abundances (and ages) obtained from
modelling the integrated spectra of early-type galaxies and bulges as
SSPs.

In Figure~\ref{fig:hbmgfe_all} I show the stellar populations of the
nuclei of early-type galaxies (data from Gonz\'alez 1993; Fisher et
al.\ 1996; Kuntschner 2000; Trager et al.\ 2003) and spiral bulges
(S0/a--Sbc; data from Proctor \& Sansom 2002).  Several results can be
read directly from this Figure.
\begin{enumerate}
  \item The nuclei of early-type galaxies and early-type spiral bulges
  have metallicities that range from just below solar to nearly five
  times the solar value.
  \item Field ellipticals, S0's in most environments, and early-type
  spiral bulges span a large range in \hbeta\ over a reasonably small
  range in \mgfe\ and therefore span a large range in SSP-equivalent
  age.  
  \item Cluster \emph{ellipticals} are typically coeval to a
  factor of $\sim2$ in age and vary mostly in metallicity (Bower,
  Lucey \& Ellis 1992; Kuntschner 2000).
  \item Ellipticals typically have $\afe>0$; \afe\ is proportional to
  velocity dispersion but apparently not to age (Trager et al.\ 2000b).
  \item S0's are slightly more ``solar'' in \afe\ than ellipticals.
  They also show a variation in \afe\ with velocity dispersion (Fisher
  et al.\ 1996), although not as strong as that in ellipticals.
  \item Bulges of early-type spirals (S0/a--Sbc) are similar to field
  ellipticals in the distribution of their ages, metallicities, and
  \afe\ ratios.
\end{enumerate}
By analyzing correlations between the derived stellar population
parameters, Trager et al.\ (2000b) found that the nuclei elliptical
galaxies (a) obey a $\log\sigma$--\afe\ relation (point 3 above, and
suggested earlier by Worthey et al.\ 1992) that may (Thomas, Maraston
\& Bender 2002) or may not (Trager et al.\ 2000b) depend on age and
(b) occupy only a thin plane in $\log t$--\z--$\log\sigma$ space (the
``$Z$-plane''), such that age and metallicity are anti-correlated with
a slope $\partial\log t/\partial\log Z=-3/2$ at fixed velocity
dispersion and metallicity increases with velocity dispersion at fixed
age.

The $\log\sigma$--\afe\ relation can be explained (Trager et al.\
2000b) as either a varying IMF with galaxy mass---larger galaxies have
``flatter'' IMFs and therefore more high mass stars and therefore more
SN II and therefore higher \afe\ and, at fixed age, higher \z\
(Worthey et al.\ 1992; Matteucci 1994)---or that smaller galaxies have
more efficient, metal-enriched winds (Vader 1986, 1987)---so that SN
II products are lost more efficiently in these objects and so have
lower \afe\ at fixed SN Ia yield and therefore lower \z\ at fixed age.
In the latter case, a high \afe\ is still required for massive
ellipticals, so a very short star formation timescale ($<0.5$ Gyr;
Matteucci 1994) or a flattened IMF would be required for all
elliptical galaxies.

A simple physical explanation for the $Z$-plane is more difficult to
find.  Multiple-burst population models seem to fit the data best,
with small (1--10\%) bursts of recent (1--3 Gyr) star formation
providing the enhanced \hbeta\ strengths and suitably increased
metallicities over a dominant (by mass) old population, but the
modelling is not unique (Trager et al.\ 2000b).  It is not yet clear
if hierarchical merging models can reproduce this plane (Kauffmann \&
Charlot 1998).  This plane has definite observational consequences.
Because age and metallicity of elliptical galaxy nuclei are correlated
at fixed velocity dispersion with a slope of 3/2 that preserves
constant broad-band optical color and metal-line indices
(\S\ref{sec:tzdegen}), the Mg--$\sigma$ relations (e.g.,
Mg$_2$--$\sigma$) are actually edge-on projections of this plane and
therefore are \emph{not} good indicators of a physically significant
mass--metallicity relation (Faber 1977; Colless et al.\ 1999).  In
fact, \emph{there is no mass--metallicity relation} in the
SSP-equivalent populations of \emph{field} elliptical galaxies;
although there \emph{is} such a relation for the roughly coeval and
presently old \emph{cluster} elliptical galaxy population.

Although multiple ages in a population are hard to recover, the
\emph{light-weighted mean metallicity} is measured quite well from an
integrated spectrum.  An Appendix in Trager et al.\ (2000b)
demonstrates the calculation based on the color-magnitude diagram of
the top of the RGB of M32 at $r\approx1.5r_e$ (Grillmair et al.\ 1996)
and extrapolated absorption-line strengths from $r\approx r_e$
(Gonz\'alez 1993).  The light-weighted metallicity from the CMD
assuming a population between 8 and 15 Gyr old is $\z=\feh=-0.25$ dex;
from the line strengths, a Worthey (1994) model (which uses the same
isochrones) gives $\z=\feh=-0.32$ dex.  A similar (though purely
theoretical) calculation for \afe\ suggests that light-weighted
abundance ratios are also measured quite reliably from integrated
spectra.

Finally, by examining other line strengths than \mgb\ and \fe, other
elemental abundances beside the rather generic \afe\ are coming under
scrutiny.  Saglia et al.\ (2002), Cenarro et al.\ (2003), and
Falc\'on-Barroso et al.\ (2003) have studied the \ion{Ca}{2} triplet
at 8600 \AA\ in early-type galaxies and bulges and found that stellar
population models (from either TMB03 or Vazdekis et al.\ 2003)
overpredict the strength of the CaT$^{\ast}$ index by as much as 1.5
\AA\ if \afe\ is set to the level suggested by \hbeta, \mgb, and \fe,
and calcium is included with the $\alpha$-elements.  This suggests
that either the IMF is significantly \emph{steeper} than previously
assumed to include more dwarf stars (cf.\ the discussion about Na D
earlier), in conflict with the $M/L$ ratios of the galaxies (Saglia et
al.\ 2002), or that calcium does not act as an $\alpha$-element in
early-type galaxies and bulges (Worthey 1998; Trager et al.\ 2000a).
This latter suggestion has recently been reinforced by Thomas,
Maraston \& Bender (2003b), who examined the Ca4227 index (centered on
the \ion{Ca}{2} $\lambda4227$ line) of ellipticals from the Lick/IDS
sample (Trager et al.\ 1998) and found that this index also required
Ca to track Fe.  More exactly, Thomas et al.\ required
$\mathrm{[\alpha/Ca]}=+0.2$ for giant ellipticals, decreasing as
velocity dispersion decreased.  Therefore, giant galaxies are Ca
underabundant.  In fact, the Ca underabundance is nearly that of the
Fe underabundance, $\mathrm{[Fe/\alpha]}=-\afe=-0.3$, so
$\mathrm{[Ca/Fe]}<+0.1$ in elliptical galaxies.

\section{Summary}

I conclude this review with an overview of the state of the art in
determining the abundances and ages of globular clusters, early-type
galaxies and the bulges of spirals from integrated-light spectra.  I
then discuss the present difficulties we have with stellar population
models and their application.  I finish with a hint of what is to come
in the immediate future.

\subsection{The current state of play}

It is now possible to determine accurately and precisely the age,
metallicity, and abundance ratio \afe\ of a simple stellar
population---one that has experienced only a single burst of star
formation---from its integrated spectrum.  We can therefore probe the
stellar populations of globular clusters, including those around
other, distant galaxies, with confidence.  In Local Group galaxies,
various researchers have used this technique to determine that the
mean metallicity of globular cluster systems correlates with the mass
of the host galaxy.  The stellar content of many clusters too distant
to resolve without adaptive optics or a large space telescope have
been probed and the rough chemical compositions characterized.
Overly-strong CN lines, apparently indicative of N enhancements, have
been detected in the Galactic, M31, and Fornax globular cluster
systems, but not in the LMC.  The stellar populations of the globular
cluster systems of early-type galaxies can now be studied in detail,
and the bimodality of globular cluster colors is now resolved into
\emph{three} populations: blue, old, metal-poor and red, old,
metal-rich, which together dominate the cluster populations, and a
frosting of red, intermediate-aged, metal-rich clusters.  The CN
anomaly is also present in these systems (Fig.~\ref{fig:exgalgcs}),
and so seems an \emph{almost} generic property of globular cluster
stellar populations---except in the LMC.

For more complex star formation histories, we can at present only
parameterize this history as three (or five) numbers without
additional information: the SSP-equivalent age, metallicity, and \afe\
(and now $\mathrm{[\alpha/Ca]}$ and $\mathrm{[N/\alpha]}$; TMB03).
However, that parameterization is a useful one, and much progress has
been made using this simplistic approach.  Field ellipticals, S0's in
many environments (except possibly in the richest clusters), and
spiral bulges span a large range in SSP-equivalent age.  Cluster
ellipticals, on the other hand, are typically coeval to within a
factor of two in age.  SSP-equivalent metallicities of the nuclei of
early-type galaxies and spiral bulges range from just below solar to
nearly five times solar.  At fixed velocity dispersion (mass), age and
metallicity are correlated in such a way as to preserve constant
broad-band optical color and metal-line strength; this makes the
interpretation of the optical color--magnitude and Mg--$\sigma$
relations of local early-type galaxies problematic.  There \emph{is} a
mass--metallicity relation for early-type galaxies \emph{at fixed
age}, which manifests itself primarily in cluster galaxies.  I note
here that line-strength gradients (Gonz\'alez 1993; Fisher et al.\
1996) suggest that early-type galaxies globally follow the same
relations, with lower metallicity and older ages (cf.\ Trager et al.\
2000a).  Early-type galaxies and spiral bulges typically have
Fe-deficient (``$\alpha$-enhanced'') stellar populations.  The value
of this deficiency, \afe, is well-correlated with velocity dispersion.
S0 galaxies are slightly more ``solar'' than giant ellipticals, where
$\langle\afe\rangle=+0.2$; bulges seem to follow the elliptical track,
although the sample of bulges available is still small.  Finally, work
is progressing on tracking the abundances of other elements, including
Ca, which seems to track Fe rather than the $\alpha$-elements in
elliptical galaxies.

\subsection{Continuing annoyances}

Of course, although we have come a long way towards precision stellar
populations, there are several significant issues still outstanding.
Here are three.
\begin{description}
\item[Oxygen abundance] The oxygen abundance controls the temperature
of the main-sequence turnoff (e.g., Salaris \& Weiss 1998), but no
direct measure of $\mathrm{[O/Fe]}$ is available in the optical.
Might the OH bands in the near-infrared be exploited for this purpose?
\item[Blue horizontal branches] The presence or not of BHB stars have
been a continuing thorn in the side of those who analyze metal-rich
stellar populations (e.g, Trager et al.\ 2000a).  They apparently
exist in dense, metal-rich clusters (Rose \& Tripicco 1986; Rich et
al.\ 1997) but the evolutionary path by which they appear in those
clusters is unknown and therefore cannot yet be included in stellar
population models.  There is also the question of the importance of
very metal-poor populations underlying the metal-rich populations that
dominate the light (Rose 1985; Maraston \& Thomas 2000).  However,
Rose (1985) showed that the strong Balmer lines in elliptical galaxies
must be dominated by the light from dwarf stars, not giants.  This
however is not true in globular clusters, as seen in
Figure~\ref{fig:mwm31gcs}.
\item[Blue straggler stars] Given their colors and luminosities, these
stars might be another significant worry, confusing age determinations
(Rose 1985; Trager et al.\ 2000a).  Again an evolutionary path for
these stars is not yet clear and therefore they have not appeared in
evolutionary syntheses.  However, Rose (1985) found little evidence of
such stars in his study of the blue line strengths of local elliptical
galaxies.
\end{description}

\subsection{Present and future directions}

The future for this subject seems bright.  Large surveys such as the
Sloan Digital Sky Survey are beginning to use this method for studying
the typical stellar populations of $10^5$ galaxies (Bernardi et al.\
2003; Eisenstein et al.\ 2003), although calibrations onto a
well-calibrated and well-modelled system are not yet available.
Two-dimensional spectroscopy using instruments like SAURON (Davies et
al.\ 2001; de Zeeuw et al.\ 2002) are now examining the spatial
distribution of the stellar populations of early-type galaxies and
bulges and combining this information with kinematics.  Spectral
synthesis of galaxies (beginning with Vazdekis 1999) and
principal-component analyses of spectra (e.g., Heavens, Jimenez \&
Lahav 2000; Reichardt, Jimenez \& Heavens 2001; Eisenstein et al.\
2003) are now coming over the horizon.  Finally, several groups are
pushing these techniques out to $z\sim1$ (Kelson et al.\ 2001; Trager,
Dressler \& Faber 2003), which will allow for the direct detection of
the evolution of old stellar populations.

\medskip
It is a pleasure to thank my collaborators, G. Worthey, S. Faber,
A. Dressler, M. Houda\-shelt, J. Dalcanton, D. Burstein, and
J. J. Gonz\'alez.  I also gratefully acknowledge very helpful
conversations with J. Brodie, A. Cole, J. van Gorkom, J. Johnson,
D. Kelson, M. Kissler-Patig, H. Kuntschner, C. Maraston, B. Poggianti,
T. Puzia, R. Schiavon, D. Thomas, and E. Tolstoy during the
preparation of this review.  The anonymous referee is thanked for a
careful reading of the manuscript and especially for pointing out the
lack of data on the inner bulge clusters of M31.  I would like to
thank the organizers, A. McWilliam and M. Rauch, for an enjoyable
meeting.  The stellar population community owes a significant debt of
gratitude to L. Robinson and J. Wampler for the development of the
Image Dissector Scanner (IDS), which revolutionized the study of
integrated spectra of galaxies.  I also would like to thank J. Nelson
and CARA for the vision to develop and build the Keck Telescopes and
J. B. Oke and J. Cohen for building the LRIS spectrograph, which
together have pushed our study of stellar populations towards the
high-redshift Universe.  This research has been supported at various
times by a Flintridge Foundation Fellowship, by a Carnegie Starr
Fellowship, by NASA through Hubble Fellowship grant HF-01125.01-99A
awarded by the Space Telescope Science Institute, which is operated by
the Association of Universities for Research in Astronomy, Inc., for
NASA under contract NAS 5-26555, and by the Kapteyn Astronomical
Institute.

\end{document}